\documentclass{elsart}
\usepackage[dvips]{graphicx}

\journal{\ \ \textbf{Physica A, Volume 268, Issues 3-4,
Pages 291-308 (1999)}.\qquad\qquad\qquad\qquad\qquad\qquad\qquad\qquad\qquad\qquad\qquad\qquad\qquad\qquad\qquad\qquad\qquad}
\input{tcilatex}
\begin{document}

\begin{frontmatter}
\title{Wetting Problem\\ For Multi-Component Fluid Mixtures}
\author[Marseille3]{Henri Gouin}\ead{henri.gouin@univ-cezanne.fr}
\quad \emph{\small and}\quad
\author[Marseille1]{Sergey
Gavrilyuk}\ead{ sergey.gavrilyuk@polytech.univ-mrs.fr}
\address[Marseille3]{Universit\'e d'Aix-Marseille \& C.N.R.S. \ U.M.R.
6181\\
  Av. Escadrille Normandie-Niemen, Box 322, 13397
Marseille Cedex 20, France.}
\address[Marseille1]{Universit\'e d'Aix-Marseille \& C.N.R.S. \ U.M.R.  6595, IUSTI,
Project SMASH\\ 5 rue E. Fermi, 13453 Marseille Cedex 13 France\\ }
\begin{abstract}
In this paper we propose an extension of the Cahn method [1] to
binary mixtures and study the problem of wetting  near a two-phase
critical point without any assumption on the form of intermolecular
potentials. A comparison between Cahn's method and later works by
Sullivan [2,3], Evans {\it et al} [4,5] is made.
 By using  an expression of
 the energy of interaction
between solid surface and liquids proposed recently by Gouin [6], we
obtain the equations of density profiles and the boundary conditions
on a solid surface. In the case of a convex free-energy, a
one-dimensional solution  of a linear problem is proposed for the
density profiles between a bulk and on a solid wall. A non-linear
model of binary  mixtures [7] extending Cahn's results for
 simple fluids is also studied. For the case of a purely attractive wall we have
established  a criterion of a first order transition in terms of the
structure of the level set  of the homogeneous part of the free
energy. Additively,  explicit expressions of density profiles  near
the wall are proposed. They allow one to consider the adsorption of
mixture components by a solid wall.
\end{abstract}
\end{frontmatter}

\section{Introduction}

In 1977, Cahn [1] gave simple illuminating arguments to describe the
interaction between solids and liquids. His model was based on a generalized
van der Waals theory of fluids treated as attracting hard spheres [7]. It
entailed assigning to the solid surface an energy that was a functional of
the liquid density "at the surface". Three hypotheses are implicit in Cahn's
picture for simple fluids: \newline
$\quad (i)$ \ In order for the liquid density to be a smooth function of the
distance from the solid surface, that surface is assumed to be flat on the
scale of molecular sizes and the correlation length is assumed to be greater
than intermolecular distances (this is the case, for example, when the
temperature $T$ is not far from the critical temperature $T_{c}$). \newline
$\quad (ii)$\ The forces between solid and liquid are of short range and can
be described simply by adding a special energy at the solid surface. \newline
$\quad (iii)$ The fluid is considered in the framework of a mean field
theory. This means, in particular that the free energy of the fluid is a
classical so-called "square-gradient functional".

After Cahn, the problem of adsorption and wetting was studied by a
statistical method by Sullivan [2,3], Evans \textit{et al}\ [4,5],
respectively for gas and binary fluid mixtures. From the point of view of
Sullivan and Evans \textit{et al}\ one may view Cahn's approach as open to
criticism for several reasons: \newline
$\quad (a)$ Cahn's treatment is based on phenomenological "square-gradient"
version of van der Waals theory, which in contrast to the approach initiated
by van Kampen [8] does not attempt to relate directly the properties of the
non-uniform fluid to the interactions occurring on a molecular level.
\newline
$\quad (b)$ The density adjacent to the wall vary strongly over the range of
intermolecular forces, consequently the gradient expansion approximation
used in deriving the square-gradient theory is no longer valid. \newline
$\quad (c)$ Cahn's theory leaves unspecified a contribution due to the
fluid-solid interfacial free energy.

Evans \textit{et al} [4,5] following Sullivan's approach [2,3] for simple
fluids consider the special case of a contact between a two-component
mixture near "the critical end point" and a wall. They used Sullivan's grand
potential to describe the solid-fluid and fluid-fluid interactions and tried
to solve directly the problem of repartition of densities in a liquid (gas).
Evans \textit{et al} obtain a coupled system of integral equations for
chemical potentials (cf. Eq. (6) in [4]). Then, to solve the system, it is
necessary to know the interaction potentials between components and between
solid wall and components: Evans \textit{et al} assume an exponential
interaction both for component-component and solid-components (as in [2,3]).
Only such a hypothesis allows one to obtain two differential equations
instead of the two integral equations (Eq. (10) in [4]). This assumption
cannot be obviously valid for large classes of mixtures. Moreover, a special
hypothesis (mixing rule) concerning interactions between components is
assumed. Then, the mixing rule and exponential dependence allow one to
obtain both the linear relation between potentials and boundary conditions
and the problem is reduced to the problem of the contact of one-component
fluid with a wall. \newline
The phenomenological "square-gradient" model is proposed in case of an
infinite non-homogeneous fluid or a fluid mixture as a small-gradient
approximation by Widom [9] and Fleming \textit{et al} [10]. The method is
extended in mean-field theory for semi-infinite media in contact with a
wall: as proved in [6], the fact that the densities are discontinuous at the
solid wall does not disqualify the procedure used by Widom and Fleming
\textit{et al} and Cahn's treatment is valid for fluids and fluid mixtures
near a critical point in contact with a wall.

\noindent In this paper, we use the expression of a surface energy. The
surface is assumed to be solid and interactions between solid and fluids are
sufficiently short-range. The contribution of fluids is represented by a
surface free energy with a density of the form \ $\ E_S (\rho_{1 S}, \rho_{2
S}) $, where \ $\rho_{1 S}$ and $\rho_{2 S} $\ are the limiting densities of
the fluid components at the surface. The expression of the surface energy
obtained in [6] is in the form:
$$
\displaystyle E_S \ = \ - \gamma_{11} \ \rho_{1 S} - \gamma_{21} \ \rho_{2
S} + {\frac{1}{2}} \ ( \gamma_{12} \ \rho^2_{1 S} + \gamma_{22} \ \rho^2_{2
S} + 2 \gamma_{32} \ \rho_{1 S} \rho_{2 S} )\, . \eqno(1)
$$
This expression represents first terms of a more complex expansion. It is an
extension with explicit calculations of the widely known expression due to
Nakanishi and Fisher [11] and examined in a review paper by de Gennes [12].
All the coefficients \ $\gamma_{ij}$ \ can be calculated explicitly after
the particular form of interaction potentials was chosen. For example, in
the case of London forces, the values of coefficients related to the
densities of the two fluids at the surface are [6]
$$
\gamma_{11} \ = \ {\frac{\mu_1 \pi}{12 \ \delta_1^2}} \ \rho_3 \ , \hskip %
0,5 cm \gamma_{21} \ = \ {\frac{\mu_2 \pi}{12 \ \delta_2^2}} \ \rho_3 \ ,
$$
$$
\gamma_{12} \ = \ {\frac{k_1 \pi}{12 \ \delta_1^2}} \ , \hskip 0,5 cm
\gamma_{22} \ = \ {\frac{k_2 \pi}{12 \ \delta_2^2}}\ ,\eqno (2) \
$$
$$
\gamma_{32} \ = \ {\frac{k_3 \pi }{{24}}} \ \ \left( {\frac{1}{\delta_1^2 }}%
+ {\frac{1}{\delta_2^2}} \right) \ ,
$$
where $\rho_3$ is the density of the solid, $\mu_i , \ i \in \lbrace 1 , 2
\rbrace$ are the coefficients associated with intermolecular potentials of
interaction between the fluids and the solid wall, $k_i , \ i \in \lbrace 1
, 2 , 3 \rbrace $ are intermolecular potentials of interaction between the
molecules of fluid $i$ and themselves or between the two fluids and \ $%
\displaystyle\delta_i = \ {\frac{1}{2}}\ (\sigma_i + \tau ),\ i \in \lbrace
1 , 2 \rbrace $ are the minimal distances between the solid and molecules of
the two species of the mixture, where $\sigma_i, \ i \in \lbrace 1 , 2
\rbrace$ is the diameter of molecule of fluid $i$ and $\tau$ for the solid.
Expression (1) allows us to estimate the influence of a solid wall on each
component of a fluid mixture. Depending on the values of coefficients \ $%
\gamma_{ij}$, one can estimate the magnitude of the attraction or repulsion
effects due to the wall. \newline
As our approach is also based on a mean-field approximation, we assume that
variations of densities near the wall take into account several molecular
ranges. Hence, it is possible to present the total free energy of the system
"fluids - wall" as the sum of a bulk free energy and a surface energy which
is an additional contribution arising from the non-uniformity of the fluid
near the wall. By using an extended variational principle, we obtain two
boundary conditions at the wall and two partial differential equations for
the density profiles of the components between a solid wall and a bulk. The
complete set of boundary conditions and equations for densities allow us to
obtain the profiles of densities in the following physical situations. The
first is the study of the linear problem associated with the equilibrium of
a two-component one-phase mixture near a critical point with a solid wall.
The second is the study of the non-linear problem of the contact between a
two-component two-phase mixture near a critical point and a wall. We get a
condition of wetting and a first order wetting transition in terms of the
level set of the homogeneous part of the free energy.

To clarify the presentation some calculations are situated in Appendices. In
Appendix 1, we present general calculations by using an extended variational
principle applied to multi-component mixtures. In Appendix 2 we give an
analytical representation of the profiles of densities connecting bulk and
solid wall for a general form of the free energy of a two-component mixture
near any critical point. These representations may be used to investigate
the adsorption of fluid components of a mixture by a solid wall.

\section{Equations of density profiles and boundary conditions: general
results}

The general form of the free energy per unit volume of the mixture is
proposed in the form [7,10,13]
$$
E = E (\displaystyle \rho_1 , \rho_2 , {\nabla \rho_1 },{\nabla \rho_2} )\, ,%
\eqno{(3)}
$$
where $\nabla$ notes the gradient operator in the physical space $\mathcal{D}
$. The associated total free energy is
$$
\mathcal{E}_{\mathcal{D}}\ =\ \int\int\int_{\mathcal{D}}\ E\ d \mathcal{D} .
$$
The wall boundary $\ \mathcal{S}\ $ of $\ \mathcal{D}\ $ is endowed with a
surface energy per unit area. The surface is solid and sharp on an atomic
scale and the interactions between surface and fluids are sufficiently short
range; the general form of the surface free energy per unit area used is
$$
E_{\mathcal{S}}\ =\ E_{\mathcal{S}}\ (\rho_{1 S},\rho_{2 S}).\eqno{(4)}
$$
Consequently, the free energy of $\mathcal{S}$ is
$$
\mathcal{E}_{\mathcal{S}} \ =\ \int\int_{\mathcal{S}} E_{\mathcal{S}}\ d%
\mathcal{S}.
$$
Then, the grand potential of the system "fluid mixture - wall" is
$$
\mathcal{E} = \ \int\int\int_{\mathcal{D}}\ E\ d \mathcal{D} + \ \int\int_{%
\mathcal{S}} E_{\mathcal{S}}\ d\mathcal{S}.
$$
The condition of extremum of the energy $\mathcal{E} $ based on hypotheses
(3) and (4) yields (see for details Appendix 1):

- \emph{Equations of two profiles of component densities}:
$$
\nabla \left ( {\frac{\partial E }{\partial \rho_\alpha}} - \mathrm{div}
\left( {\frac{\partial E}{\partial (\nabla \rho_\alpha) }} \right )  \
\right ) = 0 \quad \alpha = 1, 2 \,,\eqno {(5)}
$$
where $\displaystyle {\frac{\partial E}{\partial (\nabla \rho_\alpha) }}$ is
the vector whose components are the partial derivatives of $E$ with respect
to the components of $\nabla \rho_\alpha$ and $\ \mathrm{div \ }$ is the
divergence operator.

- \emph{Two boundary conditions at the solid wall}:
$$
\mathbf{n} \ {\frac{\partial E}{\partial (\nabla \rho_\alpha) }} + {\frac{{%
\partial E_{\mathcal{S}}}}{\partial \rho_\alpha}} = 0, \quad \alpha = 1, 2 , %
\eqno {(6)}
$$
where $\mathbf{n}$ is the external unit normal vector to $\mathcal{D}$.

\noindent Equations of equilibrium (5) are the same as in [7] given for the
one-dimensional case. Conditions (6) generalize those proposed in [1].

\section{The dynamical system associated with one-dimensional density
profiles}

In the simplest case, the surface energy per unit area is given by (1) where
the coefficients $\displaystyle\ \gamma_{ij}\ $ are expressed by means of a
mean-field approximation through the potentials of the intermolecular
interactions (see for example (2)) and the free energy per unit volume is of
the form
$$
E = U (\rho_1,\rho_2) + {\frac{1}{2}} \left ( C_1 (\nabla \rho_1)^2 + 2 D \
\nabla \rho_1\ \nabla \rho_2\ + C_2 (\nabla\rho_2)^2 \right )\,, \eqno{(7)}
$$
where $\ U (\rho_1,\rho_2)$ is the homogeneous free energy per unit volume
and $C_1, C_2, D$ are constants such that the corresponding quadratic form
is positive definite (we denote the free energy by $U$ corresponding in [7]
to $-W$).
\begin{figure}[h]
\begin{center}
\includegraphics[width=7cm]{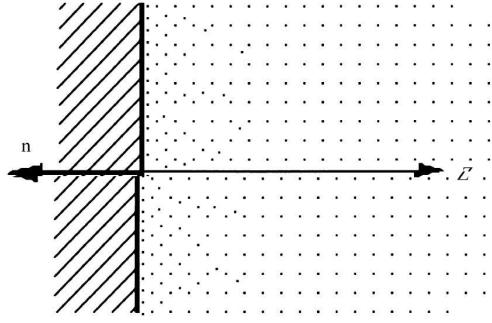}
\end{center}
\caption{\emph{One-dimensional contact of a fluid with a wall.}}
\label{fig1}
\end{figure}

\noindent Let us consider the case of a flat plate wall defined by equation $%
z = 0$ (see figure 1), where $z$ denotes the one-dimensional coordinate
orthogonal to the wall. The equations of equilibrium (6) associated with (7)
are

$$
\left \{
\begin{array}{c}
C_1 \displaystyle \ {d^2\rho_1\over dz^2} \ +\  D\ {d^2\rho_2\over dz^2}\ =\ {\partial U \over \partial\rho_1} (\rho_1,\rho_2) + e_1 \\
D \displaystyle \ {d^2\rho_1\over dz^2} \ +\  C_2 \ {d^2\rho_2\over dz^2}\ =\ {\partial U \over \partial\rho_2} (\rho_1,\rho_2) + e_2
\end{array}
\right.
\eqno{(8)}
$$

where $e_1 $ and $e_2 $ are two constants of integration.

\noindent These equations are complemented by the boundary conditions (6) at
$z = 0$. By using expression (1) of the surface energy, we get

$$
\left \{
\begin{array}{c}
C_1 \displaystyle \ {d\rho_1\over dz} \ +\  D \ {d\rho_2\over dz}\ =\ - \gamma_{11}+\gamma_{12} \rho_1+\gamma_{32} \rho_2 \\
D \displaystyle \ {d\rho_1\over dz} \ +\  C_2 \  {d\rho_2\over dz}\ =\ - \gamma_{21}+\gamma_{32} \rho_1+\gamma_{22} \rho_2
\end{array}
\right.
\eqno{(9)}
$$

We have to add the condition in the bulk ($\mathrm{at} \hskip 0,1 cm z = +
\infty $):
$$
\rho_1 = \rho_{1 \infty} , \hskip 0,2 cm \rho_2 = \rho_{2 \infty} \,. %
\eqno{(10)}
$$

\section{Linear wetting problem}

We consider the case of a one-phase mixture (liquid or gas) in contact with
a solid wall. The densities of the two-components and the temperature are
close to critical conditions. Moreover, we assume that density variations
are small enough with respect to bulk densities, i.e.
$$
{\frac{\rho _{i}-\rho _{i\infty }}{{\ \rho _{i\infty }}}}<<1,\quad i=1,2
$$%
such that we can consider a linearized problem associated with equations
(8). \noindent Let us denote
$$
\mathbf{r} =
\left ( \begin{array}{c}
\rho_1 - \rho_{1\infty} \\
\rho_2 - \rho_{2\infty}
\end{array} \right ) ,\ \ \mathbf{q}=-\left (\begin{array}{c}\gamma_{11}\\
\gamma_{21}\end{array}\right ) +\Gamma \left (\begin{array}{c}\rho_{1\infty}\\
\rho_{2\infty}\end{array}\right ) \equiv -{\mathbf{\gamma }}+\Gamma {\mathbf{\rho }_{\infty }%
},
$$%
$$
\Gamma =%
\left ( \begin{array}{cc} \gamma_{12} &
\gamma_{32} \\ \gamma_{32} & \gamma_{22} \end{array}\right ) ,\quad A=%
\left ( \begin{array}{cc} C_1 & D \\
 D & C_2
\end{array}\right ),\quad B=%
\left ( \begin{array}{cc} \displaystyle\partial^2 U\over
\displaystyle\partial \rho_1^2 & \displaystyle\partial^2 U\over
\displaystyle\partial \rho_1\partial \rho_2 \\
\displaystyle\partial^2 U\over \displaystyle\partial \rho_1 \partial
\rho_2 & \displaystyle\partial^2 U\over \displaystyle\partial
\rho_2^2 \end{array}\right ).\eqno(11)
$$%
The matrix $B$ is calculated in the bulk $(\rho _{1\infty },\rho _{2\infty
}) $. Taking into account the definitions (11), we get the linearized
problem associated with equations (8)-(10) in the form:
$$
A{\frac{d^{2}\mathbf{r}}{dz^{2}}}=B\mathbf{r\ \ \ \ \ \ \ \ \ \ \ \ \ \ \ \
\ \ \ }\eqno(12)
$$%
$$
A{\frac{d\mathbf{r}}{dz}}=\mathbf{q}+\Gamma \mathbf{r}\quad \mathrm{at}\quad
z=0 \eqno(13)
$$%
$$
\mathbf{r}=0\quad \mathrm{at}\quad z=+\infty .\ \ \ \ \ \eqno(14)
$$%
The stability of the thermodynamic state of the bulk requires that the
symmetric matrix $B$ is also positive definite.

\noindent Let $\chi_i^2$, $\mathbf{h}_i$ be the eigenvalues and the
eigenvectors of the equation
$$
\left(B - \chi_i^2 A \right)\mathbf{h}_i =\mathbf{0} .
$$
Since $B$ and $A$ are symmetric and positive definite, $\chi_i^2$ are
positive. We can always suppose that\footnote{$\mathbf{V} N \mathbf{W}
\equiv \mathbf{V}\otimes(N \mathbf{W})$ denotes the bilinear form of
vectors $\mathbf{V}$ and $\mathbf{W}$ with respect to matrix $N$; the
bilinear form is symmetric when matrix $N$ is symmetric.}
$$
\mathbf{h}_i A \mathbf{h}_j = 0, \quad i\neq j\,.
$$
The solution of (12) satisfying the condition (14) is in the form
$$
\mathbf{r} = \sum_{i =1}^{2}b_i \mathbf{h}_i\exp({-\chi_i z})\quad \mathrm{%
where} \quad \chi_i > 0 \,.\eqno(15)
$$
Substituting expression (15) into condition (13), we get a linear system of
algebraic equations for the unknown coefficients $b_i$
$$
\sum_{i =1}^{2}b_i (\Gamma + \chi_i A)\mathbf{h}_i = - \mathbf{q}
$$
which defines a unique solution $b_i$ if $det (\mathbf{h}_i\Gamma \mathbf{h}%
_j + \chi_i \mathbf{h}_i A \mathbf{h}_j ) \neq 0 $. In particular, if $%
\Gamma $ is negligible (we assume that the wall is purely attractive), we
get $\ \mathbf{q} = - \ {\mathbf{\gamma}} $ (see (11)) and
$$
b_i = {\frac{{\mathbf{\gamma}} \ \mathbf{h}_i }{{\chi_i(\mathbf{h}_i A
\mathbf{h}_i )} }}\,.
$$
In such a case, the solutions satisfy conditions (13) and the density
profiles fulfil the solution of linearized problem. Equations (15) yield
different forms of density profiles. Depending on wall conditions, we may
obtain both monotonic and non-monotonic profiles. This is similar to results
of [7] in the non-linear case without a solid wall. In figure 2, we
represent the different density profiles for each component of the mixture.
We note that only one extremum point may appear for each density profile.
This result is different from the results of Evans \textit{et al} where
density profiles are essentially monotonic.
\begin{figure}[h]
\begin{center}
\includegraphics[width=13cm]{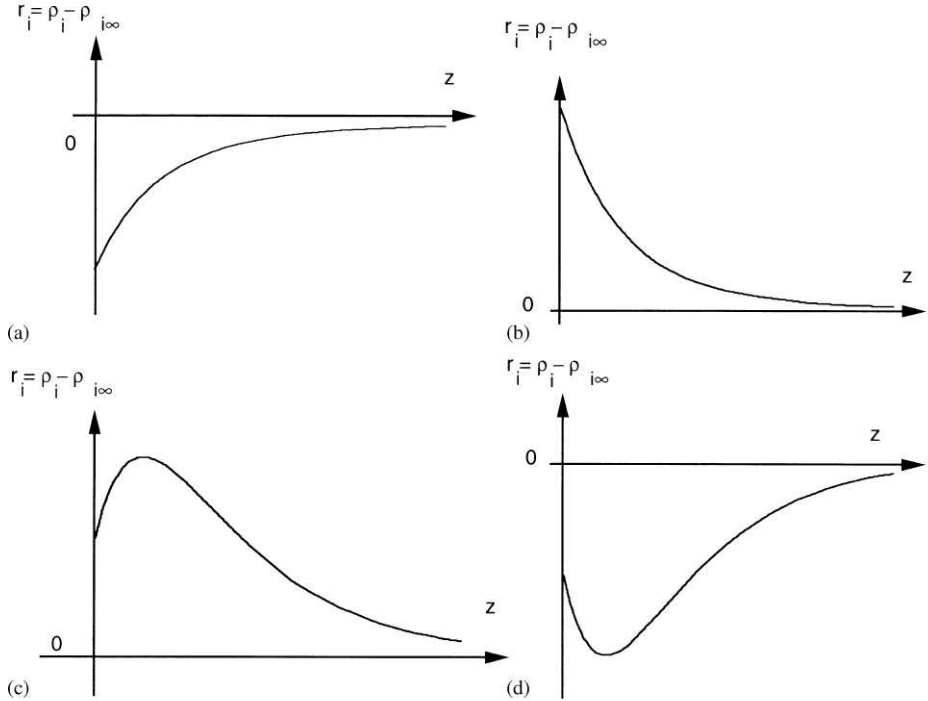}
\end{center}
\caption{\emph{\ In (a) and (b) the profiles of densities for a component
are monotonic. In (c) and (d) the profiles admit an extremum of density for
a component. The extremum is not on the wall but at some distance into the
bulk measured on a molecular scale. }}
\label{fig1}
\end{figure}

\section{Wetting problem near a critical point for a two-component mixture}

\subsection{The dynamical system}

In a two-phase region near a critical point at a given temperature $T$, the
expression of the free energy per unit volume $U$ associated with a phase
equilibrium is of the form [7]
$$
U = a_0^2 \Big( (b_0^2 x^2 + y)^2 + (y+t) ^2 \Big) . \eqno(16)
$$
The parameter $t$ is an independently varied field characterizing the
"distance" from the critical point $(\rho_{1c},\rho_{2c})$, $a_0$ and $b_0$
are functions of the temperature. The variables $x$ and $y$ are defined
through the transformation
$$
\mathbf{r}= \Delta \mathbf{R},\, \quad \mathbf{R} =\left ( \begin{array}{c}x \\ y\end{array} \right ),
\quad \Delta =\left ( \begin{array}{cc}a & b \\ c & d \end{array}\right ),\quad \mathbf{r} = {\mathbf{%
\rho}}- {\mathbf{\rho}}_c,
$$
$$
{\mathbf{\rho}}= \left ( \begin{array}{c}\rho_{1} \\\rho_{2} \end{array}\right ) , \quad {\mathbf{\rho}}_c =
\left ( \begin{array}{c}\rho_{1c} \\
\rho_{2c}\end{array}\right ) \,. \eqno(17)
$$
The scalars $a,b,c,d$ associated with the physical properties of the mixture
near the critical point depend on the temperature $T$. The constants of
integration $e_1=0$ and $e_2=t$ are already incorporated in $U$.  With Eq.
(17), the system (8) can be rewritten in the form
$$
\Delta^{\star } A\Delta {\frac{d^2 \mathbf{R}}{dz^2}} = \nabla _{\mathbf{R}}
U , \eqno(18)
$$
where $\Delta^{\star}$ denotes the transpose matrix and $\nabla _{\mathbf{R}%
} U$ means the gradient with respect to $\mathbf{R}$. \newline
Following Rowlinson and Widom [7] we denote by
$$
M = \Delta^{\star} A\Delta =
\left ( \begin{array}{cc}m_{xx} & m_{xy}\\
 m_{xy} & m_{yy}
\end{array}\right ). \eqno(19)
$$
Obviously, if $A$ is positive definite, $M$ is also positive definite, i.e. $%
m_{xx} > 0 $, $m_{xx} m_{yy} - m_{xy}^2 > 0 $. The boundary conditions (9)
at the wall are
$$
M {\frac{d \mathbf{R}}{dz}} = \mathbf{g} + G \ \mathbf{R}, \quad \mathrm{%
where}\quad \mathbf{g} = \Delta^{\star}(- {\mathbf{\gamma}} + \Gamma {%
\mathbf{\rho}}_c) \quad \mathrm{and}\quad G = \Delta^{\star} \Gamma\Delta
\,. \eqno(20)
$$
In the following, we choose in Eq. (16) $a_0 = 1/\sqrt{2}$ and $b_0 = 1$ (to
do this, we have only to change the values of coefficients of the matrix $%
\Delta $ defined by (17)). Hence, $\quad U = \displaystyle {\frac{1}{2}} %
\Big ( (x^2 + y)^2 + (y+t) ^2 \Big) $.

\noindent The system (18)-(19) yields
$$
\left \{
\begin{array}{c}
\displaystyle m_{xx}{d^2 x\over dz^2} + m_{xy}{d^2 y\over dz^2} = 2x(x^2 + y) \\
\displaystyle m_{xy}{d^2 x\over dz^2} + m_{yy}{d^2 y\over dz^2} = x^2 + 2y + t
\end{array}
\right.
\eqno{(21)}
$$
System (21) admits the first integral
$$
{\frac{1}{2 }}\ m_{xx} ({\frac{dx}{dz}})^2 + \ m_{xy} ({\frac{dx}{dz}}) ({%
\frac{dy}{dz}}) + {\frac{1}{2 }}\ m_{yy} ({\frac{dy}{dz}})^2 - U(x,y) = 0 . %
\eqno(22)
$$
This integral is similar to the integral of energy for mechanical problems.

\noindent Substitution of boundary conditions (20) into the relation (22)
yields necessary conditions for $x, y$ at the solid wall. For simplicity, we
consider only the case of an attractive wall ($G$ is then negligeable).
Conditions (20) yield
$$
M {\frac{d\mathbf{R}}{dz}} = \mathbf{g} . \eqno (23)
$$
In fact, it is natural to expect that the results we obtain in the case of
an attractive wall are closely similar to the results associated with the
most general case. Relations (22) and (23) yield
$$
U(x,y) = k^2 ,
$$
where $k^2 = \mathbf{g} M^{-1} \mathbf{g} $. Then, discussion of the wetting
of a fluid mixture with a solid wall arises naturally from the drawing of
the level curves of U(x,y) as a function of the parameter $t$.

\subsection{ Connection between the dynamical system and Young's conditions}

For a solid wall in contact with phases $\alpha$ and $\beta$, the contact
angle $\theta$ is defined with the help of surface free energies $\sigma$
along the solid surface (Young's conditions)
$$
\sigma_{\alpha \beta} \ cos \theta = \sigma_{\alpha S} \ - \ \sigma_{\beta
S} \,.\eqno (24)
$$
where the different subscripts designate phases adjoining the surface or
interface. No value of $\theta$ satisfies Eq. (24) unless
$$
\sigma_{\alpha \beta} > \vert \ \sigma_{\alpha S} \ - \ \sigma_{\beta S}\
\vert \,.\eqno (25)
$$
If the inequality (25) is not satisfied, one of the fluid phases completely
wets the solid and there is no contact between solid and other fluid phase.
In fact, the forbidden surface is replaced by a layer of the wetting phase
and the surface free energy becomes the sum of two surfaces' free energies
of the layer
$$
\sigma_{\alpha S} = \sigma_{\alpha \beta} \ + \ \sigma_{\beta S} . \eqno  %
(26)
$$
Condition (26) corresponds to the perfect wetting with the solid. The
surface energies can be calculated by the formulas
$$
\sigma_{\alpha \beta} = \int_{\ -\infty}^{\ \ \ \ +\infty} (K + U) dz ,
\quad \sigma_{\alpha S} = \int _{ \ 0}^{ \ \ \ \ +\infty} (K + U) dz, \quad
\sigma_{\beta S} = \int _{ \ 0}^{\ \ \ \ +\infty} (K + U) dz ,
$$
where $\displaystyle K = {\frac{1}{2 }}\ m_{xx} ({\frac{dx}{dz}})^2 + \
m_{xy} ({\frac{dx}{dz}}) ({\frac{dy}{dz}}) + {\frac{1}{2 }}\ m_{yy} ({\frac{%
dy}{dz}})^2 \ $ and integrals are taken on different paths connecting phase $%
\alpha$ and phase $\beta$ or a phase and the wall [7].

\noindent Let us note $\displaystyle \ (ds)^2 = {\frac{1}{2 }}\ m_{xx}
(dx)^2 + \ m_{xy} dx dy + {\frac{1}{2 }}\ m_{yy} (dy)^2 $.

\noindent From the first integral (22), we get
$$
\sigma _{\alpha \beta } =\int_{\ \ \ \ \ \ (x_{\alpha },y_{\alpha })}^{\ \ \
\ \ \ \ \ \ (x_{\beta },y_{\beta })}(2U(x,y))^{\frac{1}{2}%
}ds,\quad\quad\quad \sigma _{\alpha S}=\int_{\ \ \ \ \ \ \ \ \ \
(x_{M_{\alpha }}, y_{M_{\alpha }})}^{\ \ \ \ \ \ \ \ \ (x_{\alpha },
y_{\alpha })}(2U(x,y))^{\frac{1}{2}}ds,
$$
$$
\sigma _{\beta S} =\int_{\ \ \ \ \ \ \ \ \ \ (x_{M_{\beta }}, y_{M_{\beta
}})}^{\ \ \ \ \ \ \ \ \ (x_{\beta }, y_{\beta })}(2U(x,y))^{\frac{1}{2}}ds
\,.\eqno (27)
$$
The integrals (27) are calculated on the paths associated with system (21)
and the boundary conditions on the wall
$$
x=x_{M_{\alpha }},\quad y=y_{M_{\alpha }}\quad \mathrm{or}\quad
x=x_{M_{\beta }},\quad y=y_{M_{\beta }},
$$
and in the bulks
$$
x=x_{\alpha },\quad y=y_{\alpha }\quad \mathrm{or}\quad x=x_{\beta },\quad
y=y_{\beta }.
$$

\subsection{Discussion of the wetting}

For a solid wall the value of $k^2$ is given. Hence, the discussion depends
on the relative value of parameter $t$.

(a) $t > 0$ and large enough.

\noindent In this case we are far enough from the critical conditions. In
figure 3a the phases are in points $A(x_\alpha,y_\alpha)$ and $%
B(x_\beta,y_\beta)$. One obtains easily that $\ x_\alpha = -\sqrt t,\
y_\alpha = -t,\ x_\beta =\sqrt t,\ y_\beta = -t $. The points $M_\alpha$ and
$M_\beta$ belong to two different connected components of the level set $U =
k^2$. In the vicinity of $A$ (or $B$), the energy $U$ is a convex function
of $x,y$ and as in Section 4, it is possible to find the profiles of
densities connecting $A$ and $M_\alpha$ or $B$ and $M_\beta$, respectively.
The integrals (27) are positive and $\sigma_{\alpha,\beta}$ is large with
respect to $\sigma_{\alpha,S}$ and $\sigma_{\beta,S}$. Then the relations
$$
\sigma_{\alpha,S} \le \sigma_{\alpha,\beta} + \sigma_{\beta,S} \quad \mathrm{%
and} \quad \sigma_{\beta,S} \le \sigma_{\alpha,\beta} + \sigma_{\alpha,S} %
\eqno (28)
$$
hold and we are in the case of partial wetting with $\theta \ne 0$.

(b) $t > 0$ and small enough.

\noindent This case corresponds to phases close enough to the critical point
(see figure 3b). The level set $U = k^2$ consists only of one connected
component containing the points $M_\alpha$ and $M_\beta$. The phases are at
the points $A(x_\alpha,y_\alpha)$ and $B(x_\beta,y_\beta)$. They are very
close with respect to the distance to the level curve. The superficial
tension $\sigma_{\alpha,\beta} $ is small with respect to the free energies $%
\sigma_{\alpha,S}$ and $\sigma_{\beta,S}$. The values of $\sigma_{\alpha,S}$
and $\sigma_{\beta,S} $ are in general different and one of the two
relations (28) is not satisfied. We are in the case where one of the two
phases wets completely the solid wall. No contact appears between the other
phase and the solid. For exemple, if relation (26) is satisfied, the phase $%
\beta$ wets completely the wall.
\begin{figure}[h]
\begin{center}
\includegraphics[width=14.5cm]{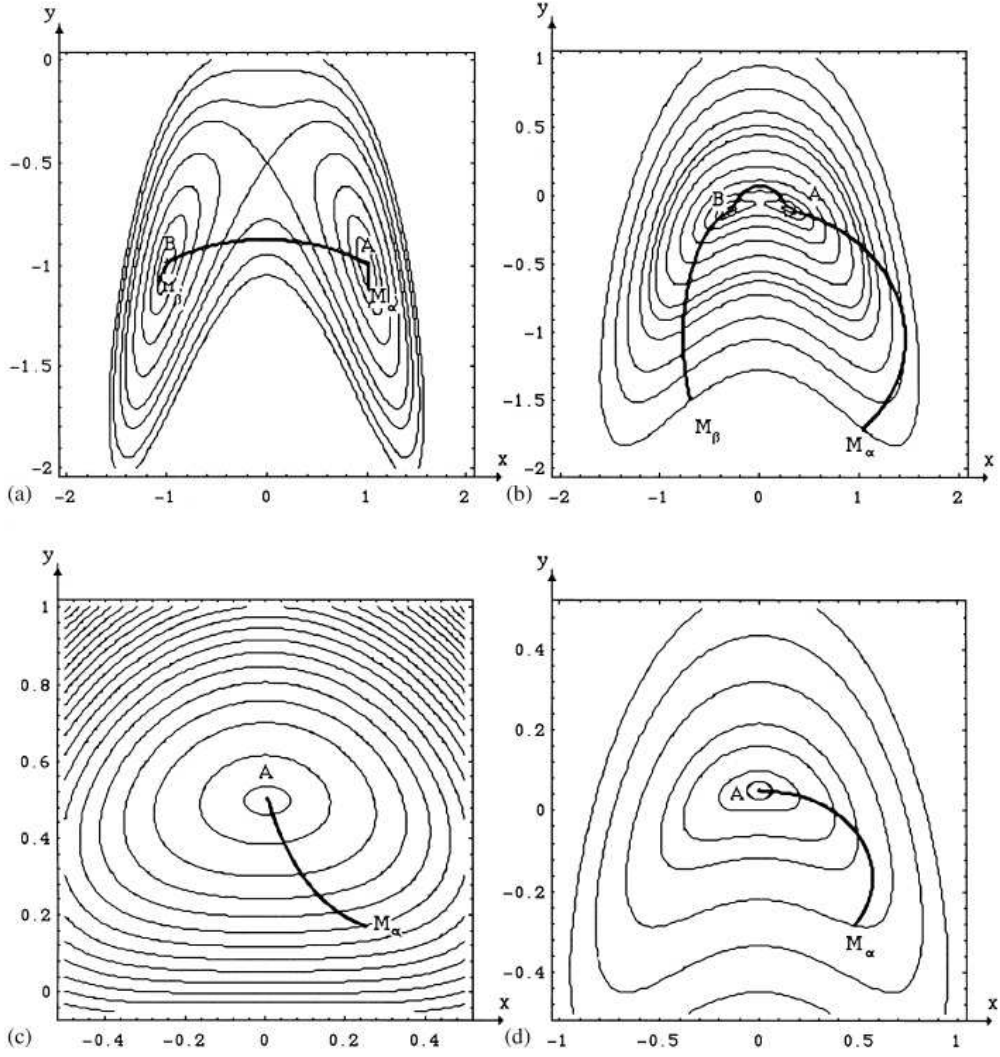}
\end{center}
\caption{\emph{Drawing of level curves for the free energy U for different
values of t. Points A (and B) correspond to the bulks. The bold curves are
paths connecting a phase and the solid wall or two phases. The interfacial
tension is calculated along these paths. (a) is the case of partial wetting
with a non-zero Young angle. The other figures are associated with different
cases of total wetting of one phase: in (b) with two phases and in (c) and
(d) with one phase. }}
\label{fig1}
\end{figure}

(c) $t < 0$

The mixture has only one phase at the point $A(0, - t/2)$, which is the only
singular point of the system (21). The energy $U$ attains a minimum at the
point $A$ (we note that for $t>0$ this point corresponds to a saddle point,
which is not associated with a bulk phase). The free energy $U$ of the
mixture is convex at the vicinity of $A$ (figures 3c and 3d). If $t$ is
small enough, the linear solution for the profiles of densities obtained in
Section 4 can be used. When $t$ is large enough, the solution for the
profiles of densities can be calculated analogously as in the Appendix 2.

\subsection{Some remarks on the profiles of densities}

The system (21) yields
$$
M\ \ {\frac{d^2\mathbf{R}}{dz^2}}\ =\ \left(%
\begin{array}{c}\displaystyle 2x(x^2+y) \\ \displaystyle x^2+2y+t \end{array} \right) \eqno{(29)}
$$
and admits the first integral (22):
$$
\ {\frac{d\mathbf{R}}{dz}}\ M\ {\frac{d\mathbf{R}}{dz}}\ -\
(x^2+y)^2-(y+t)^2 = 0 .\eqno{(30)}
$$
When the densities \ $x,y\ $ are far from critical conditions, $t\ $ is
negligeable with respect to \ $x\ $ and \ $y\ $ and (30) reads
$$
\ {\frac{d\mathbf{R}}{dz}}\ M\ {\frac{d\mathbf{R}}{dz}}\ -\ (x^2+y)^2-y^2 =
0 .\eqno{(31)}
$$
Let us denote \ $V(z)\ =\ \mathbf{R}M\mathbf{R}.\ $ Then, by using (29) and
(31), we get \ $\ \displaystyle
{\frac{d^2V}{dz^2}}\ =\ 3x^4+5x^2y+3y^2\ $. The right-hand side is a
positive definite quadratic form, which implies that
$$
\ {\frac{d^2V}{dz^2}}\ >\ 0\,.
$$
Hence,
$$
V(z)\ \ge\ V(0)+V^{\prime }(0)z \,.
$$
If \ $\displaystyle \ {\frac{dV}{dz}}\Bigg\vert_{ z=0}\ >\ 0\ $ , it follows
from here that $V\rightarrow \infty$ as $z\rightarrow\infty$. The level
curves of \ $V\ $ are represented in figure 4a. Hence, \ $x\ $ or \ $y\ $
must be an increasing function of \ $z\ $ near the solid wall. For example,
let \ $x\ $ be an increasing function of \ $z\ $ when \ $z\ $ is small
enough. Due to the fact that $x\rightarrow\ \pm\ \sqrt t$ as $z\rightarrow\
+\infty$ and \ $t\ $ is small with respect to \ $x(0), \ $ the
representation of \ $x\ $ as a function of \ $z\ $ has the form shown in
figure 4b. Hence, the function \ $x(z)\ $ is non-monotonic. In this case we
may have also non-monotonic profiles of densities unlike in the treatment of
Evans et al. Then, construction of an analytical solution may be done
according to the algorithm proposed in Appendix 2.
\begin{figure}[h]
\begin{center}
\includegraphics[width=12cm]{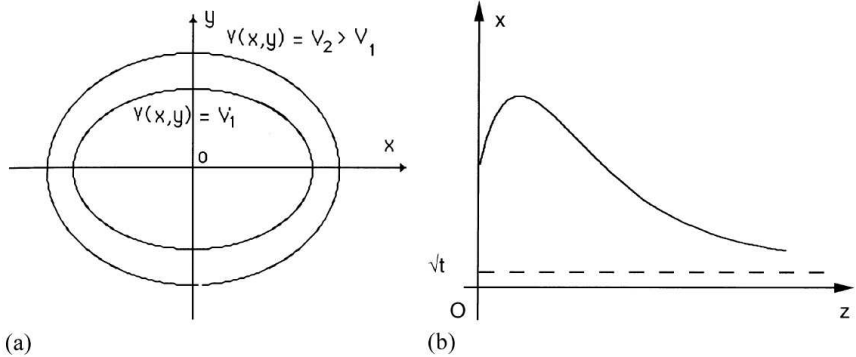}
\end{center}
\caption{\emph{(a) represents level sets of V. Since x or y must be an
increasing function of z near the wall, due to the limit conditions at
infinity, x or y is a non-monotonic function of z. (b) is a representation
of such a function. }}
\label{fig1}
\end{figure}

\section{Conclusion}

Near critical conditions, by using a variational approach, we have obtained
for an isothermal binary mixture in contact with a solid wall equations of
equilibrium and boundary conditions which generalize those obtained by Cahn.
With limit conditions in the bulk, they form a closed boundary value
problem. When the free energy of the mixture is a quadratic form with
respect to the densities of components and their gradients, we get explicit
profiles of the densities in the one-dimensional case. \newline
In the case of a purely attractive wall we have also established a criterion
of a first order transition, when a contact angle against a solid wall
becomes zero. This criterion is formulated in terms of the level set of the
function $U(x,y)$: $U(x,y) = k^2$, where $k^2$ depends on the boundary
conditions. If the level set is a connected set, two multi-component layers
exist: one layer with ordinary adsorption and the second one in contact with
the wetting layer. If the level set is disconnected we have partial wetting.
We have also shown that the profiles of density are typically non-monotonic.
This is in agreement with Rowlinson and Widom [7] where infinite two-phase
two-component mixtures where considered.

\textbf{Acknowledgements}

We have greatly benefited from comments and advises of Professor Benjamin
Widom. \bigskip

\textbf{Appendix 1.\quad Calculus of variations for fluid mixtures }

\medskip We study a two-fluid equilibrium, but the method can be extended to
any number of components. The position of a two-fluid mixture is associated
with two applications
$$
\mathbf{x}\ =\ {\mathbf{\phi}}_\alpha\ (\mathbf{X}_\alpha) , \hskip 0,2 cm
\alpha = 1, 2 ,
$$
where $\ \mathbf{X}_\alpha\ $ denote the Lagrangian coordinates belonging to
a reference space $\ \mathcal{D_\alpha\ }$ associated with the $\ \alpha$th
component and $\ \mathbf{x}\ $ denotes the Eulerian coordinates in the
physical space $\ \mathcal{D}$ [13]. The virtual motions of particles are
deduced from the relation
$$
\displaystyle \ \mathbf{x} = \mathbf{\Phi}_\alpha (\mathbf{X}_\alpha ,
\varepsilon _\alpha) , \ \mathbf{\Phi}_\alpha (\mathbf{X}_\alpha , 0 )\ = {%
\mathbf{\phi}}_\alpha (\mathbf{X}_\alpha) \,.
$$
Here $\ \displaystyle
\varepsilon _\alpha ,\ \alpha=1,2\ $ are small parameters defined in a
neighbourhood of zero. Virtual displacement $\displaystyle\ {\mathbf{\zeta}}%
_\alpha $ are defined by [13,14]
$$
{\mathbf{\zeta}}_\alpha \ = {\frac{\partial\mathbf{\Phi}_\alpha }{
\partial\varepsilon_\alpha}}(\mathbf{X}_\alpha, \varepsilon_\alpha)_{%
\displaystyle \vert_{\varepsilon_\alpha=0}}, \quad \alpha = 1,2 \,. %
\eqno{(A1)}
$$
At the solid boundary, the virtual displacement ${\mathbf{\zeta}}_\alpha $
is subject to the conditions
$$
\mathbf{n} \, {\mathbf{\zeta}}_\alpha = 0 , \ \alpha = 1,2 \,,\eqno (A2)
$$
where $\mathbf{n}$ is the unit normal vector to the boundary.

\noindent Eulerian variations of densities are defined by
$$
\delta_\alpha \rho_\alpha = {\frac{d }{{d \varepsilon_\alpha }}} \rho_\alpha
(\mathbf{x}, \varepsilon_\alpha)_{ \displaystyle \vert_{\varepsilon_%
\alpha=0}} \quad \mathrm{and} \quad \delta_\alpha \rho_\beta = 0 ,\quad
\beta \neq \alpha , \quad \alpha, \beta = 1,2 \,.\eqno (A3)
$$
The variations (A3) are related to the virtual displacements (A1) by the
formulae [14,15]
$$
\delta_\alpha \rho_\alpha =- \mathrm{div}(\rho_\alpha \ {\mathbf{\zeta}}%
_\alpha)\ ,\ \ \alpha = 1,2 .\eqno (A4)
$$
The variations of the volume free energy are
$$
\delta_\alpha \mathcal{E}_{\mathcal{D}}\ =\ \int\int\int_{\mathcal{D}}
\delta_\alpha E d\mathcal{D} ,
$$
where
$$
\displaystyle\ \delta_\alpha E\ =\ {\frac{\partial E}{\partial \rho_\alpha}}%
\delta_\alpha\rho_\alpha \ +\ \mathbf{f}_\alpha\ \delta_\alpha {\nabla
\rho_\alpha } \quad \mathrm{with} \quad \displaystyle\ \mathbf{f}_\alpha\ = {%
\frac{\partial E}{\partial (\nabla \rho_\alpha) }}\,.
$$
Since
$$
\displaystyle\ \delta_\alpha \nabla \rho_\alpha \ = \ {\nabla( \delta_\alpha
\rho_\alpha)}
$$
we get
$$
\displaystyle\delta_\alpha \mathcal{E}_{\mathcal{D}}\ = \ \int\int\int_{%
\mathcal{D}} \left(\displaystyle   {\frac{{\delta E }}{{\delta \rho_\alpha }
}}\, \delta_\alpha \rho_\alpha + \hbox{ div }  (\mathbf{f}_\alpha\,
\delta_\alpha \rho_\alpha) \right) d\mathcal{D}, \eqno{(A5)}
$$
where the variational derivative $\displaystyle {\frac{\delta E}{%
\delta\rho_\alpha}}$ is defined by
$$
\displaystyle\ {\frac{\delta E}{\delta\rho_\alpha}}\ \equiv \ \ {\frac{%
\partial E}{\partial \rho_\alpha}}\ -\ \hbox{div } \mathbf{f}_\alpha .
$$
From relations (A2), (A4) and (A5) we obtain
$$
\displaystyle\delta_\alpha \mathcal{E}_{\mathcal{D}}\ =\ \int\int\int_{%
\mathcal{D}} \left( \displaystyle  \rho_\alpha \nabla \left({\frac{{\delta E
}}{{\ \delta \rho_\alpha } }} \right) {\mathbf{\zeta}}_\alpha - \hbox{
div } \left( \rho_\alpha {\frac{\delta E }{{\ \delta \rho_\alpha }}} \, {%
\mathbf{\zeta}}_\alpha + \mathbf{f}_\alpha \hbox{ div } ( \rho_\alpha {%
\mathbf{\zeta}}_\alpha ) \right ) \right) \ d \mathcal{D}
$$
$$
=\ \int\int\int_{\mathcal{D}} \displaystyle\ \rho_\alpha {\nabla %
\displaystyle  \left({\frac{{\delta E }}{{\ \delta \rho_\alpha } }} \right) {%
\mathbf{\zeta}}_\alpha \ d \mathcal{D}\displaystyle
- \int\int_{\mathcal{S}} {\displaystyle  \mathbf{n} \ \mathbf{f}_\alpha
\hbox{
div } ( \rho_\alpha {\mathbf{\zeta}}_\alpha }) d\mathcal{S}}\,.
$$
The variations of the surface free energy are
$$
\delta_\alpha \mathcal{E}_{\mathcal{S}} = \ \int\int_{\mathcal{S}} \ {\frac{%
\partial E_{\mathcal{S}} }{ \partial \rho_\alpha}}\, \delta_\alpha
\rho_\alpha d\mathcal{S} = \ - \int\int_{\mathcal{S}} \ {\frac{\partial E_{%
\mathcal{S}} }{\partial \rho_\alpha}}\, \mathrm{div} (\rho_\alpha {\mathbf{%
\zeta}}_\alpha) d\mathcal{S} \,.
$$
The grand potential of the system is $\mathcal{E} = \mathcal{E}_{\mathcal{D}%
} + \mathcal{E}_{\mathcal{S}} $ and its $\alpha$th variation is given by the
formula
$$
\delta_\alpha \mathcal{E} = \ \int\int\int_{\mathcal{D}} \displaystyle\
\rho_\alpha {\nablaÊ \displaystyle  \left({\frac{{\delta E }}{{\ \delta
\rho_\alpha } }} \right) {\mathbf{\zeta}}_\alpha \ d \mathcal{D}%
\displaystyle
\ - \int\int_{\mathcal{S}} {\displaystyle \left( \mathbf{n} \ \mathbf{f}%
_\alpha + {\frac{\partial E_{\mathcal{S}} }{\partial \rho_\alpha}}\right) %
\hbox{ div } (\rho_\alpha {\mathbf{\zeta}}_\alpha }) d\mathcal{S}} \,.
$$
Denoting $c_\alpha = {\displaystyle   \mathbf{n}\, \mathbf{f}_\alpha + {%
\frac{\partial E_{\mathcal{S}} }{\partial \rho_\alpha}}}$, we obtain
$$
\int\int_{\mathcal{S}} c_\alpha \mathrm{div } (\rho_\alpha {\mathbf{\zeta}}%
_\alpha) d\mathcal{S} = \int\int_{\mathcal{S}} \left( c_\alpha \ \mathrm{div}%
_{\mathcal{S}} (\rho_\alpha {\mathbf{\zeta}}_{\alpha}) + c_\alpha \left(%
\mathbf{n} \frac{\partial (\rho_\alpha {\mathbf{\zeta}}_\alpha)} {\partial
\mathbf{x}}\mathbf{n} \right)  \right) d\mathcal{S} \,,
$$
where $\mathrm{div}_{\mathcal{S}}$ denotes the surface divergence. Denoting
by ${\nabla}_{\mathcal{S}}$ the tangential gradient to $\mathcal{S}$, we get
finally
$$
\delta_\alpha \mathcal{E} = \int\int\int_{\mathcal{D}} \displaystyle
\rho_\alpha \nabla \displaystyle   \left({\frac{{\delta E }}{{\ \delta
\rho_\alpha } }} \right) {\mathbf{\zeta}}_\alpha d \mathcal{D}\displaystyle
+ \int\int_{\mathcal{S}} \left( \rho_\alpha \nabla_{\mathcal{S}} (c_\alpha) {%
\mathbf{\zeta}}_\alpha - c_\alpha \rho_\alpha \left(\mathbf{n} \frac{%
\partial \mathbf{\zeta}_\alpha} {\partial \mathbf{x}}\mathbf{n}\right)
\right) d\mathcal{S}
$$
Consequently, the equations of equilibrium are
$$
\nabla \displaystyle  \left( {\frac{ {\delta E }}{{\ \delta \rho_\alpha} }}
\right) =\ 0 , \quad \alpha = 1,2 \,,\eqno{(A7)}
$$
or
$$
\nabla \left ( {\frac{\partial E }{\partial \rho_\alpha}} - \mathrm{div}
\left( {\frac{\partial E}{\partial (\nabla \rho_\alpha) }} \right ) \ \right
) = 0 \quad \alpha = 1, 2
$$
and the boundary conditions are
$$
c_\alpha =0 , \quad \mathrm{and } \quad \nabla_{\mathcal{S}} (c_\alpha) = 0
, \quad \alpha = 1, 2 .
$$
Due to the fact that $\nabla_{\mathcal{S}} (c_\alpha) = 0$ is a direct
consequence of relation $c_\alpha =0$ on the surface, the only effective
boundary conditions are $c_\alpha =0$ , i.e.
$$
\mathbf{n} \ {\frac{\partial E}{\partial (\nabla \rho_\alpha) }} + {\frac{{%
\partial E_{\mathcal{S}}}}{\partial \rho_\alpha}} = 0, \quad \alpha = 1, 2
\,.
$$
\medskip

\textbf{Appendix 2.\quad Analytical representation of the profiles of
densities of a two-component mixture for the wetting problem near a critical
point }

\medskip

Our purpose is to express analytically the profiles of densities of a
two-phase mixture in contact with a solid wall. The free energy is given by
(16) and the dynamical system for the profiles are given by
$$
\left \{
\begin{array}{c}\displaystyle m_{xx}{d^2 x\over dz^2} +
m_{xy}{d^2 y\over dz^2} = 2x(x^2 + y) \\ \displaystyle m_{xy}{d^2 x\over dz^2} +
m_{yy}{d^2 y\over dz^2} = x^2 + 2y + t \end{array} \right. \eqno (A8)
$$
We notice that the matrix
$$
{\frac{\partial^2 U }{\partial\mathbf{R}^2}} =
\left(\begin{array}{cc}6x^2 + 2y & 2x\\  2x & 2\end{array}\right)
$$
is positive definite in the bulk phases $(\pm x_{00}, y_{00})$ where $%
y_{00}= - x_{00}^2$ and $t= x_{00}^2 \neq 0$. As in the Section 4, we can
determine the positive eigenvalues $\chi^2$ defined from the equation
$$
\mathrm{det} \left({\frac{\partial^2 U }{\partial\mathbf{R}^2}}- \chi^2
M\right) =0 ,
$$
where all the matrix coefficients are calculated in the bulk. We obtain
$$
\chi^4(m_{xx}m_{yy} -m_{xy}^2) + \chi^2(4x_{00}m_{xy}- 4x_{00}^2m_{yy} -
2m_{xx}) + 4x_{00}^2 = 0 .
$$
For $\vert x_{00}\vert$ small enough, we get the two eigenvalues in the form
$$
\lambda^2 \equiv \chi_1^2 = {\frac{2 x_{00}^2}{m_{xx}}} + \mathcal{O}(\vert
x_{00}\vert^3) \eqno(A9)
$$
and
$$
\mu^2 \equiv \chi_2^2 = \mu_0^2 - {\frac{4m_{xy}}{D^2}} x_{00}+ \mathcal{O}%
(\vert x_{00}\vert^2) ,\quad \mathrm{where}\quad \mu_0^2 = 2m_{xx}/ D^2 ,
$$
$$
D^2 = m_{xx}m_{yy} -m_{xy}^2 \,.\eqno(A10)
$$

\noindent We are looking for the solution of the system (A8) which goes to
the equilibrium states $(\pm \sqrt{t},-t)$, $t>0$ at $z=+\infty $ in the
following form
$$
\left\{
\begin{array}{c} \displaystyle x = \sum_{k,m \ge 0} x_{km} \exp \big(-(\lambda k + \mu
m)z \big) \\ \displaystyle y = \sum_{k,m \ge 0} y_{km} \exp \big(-(\lambda k + \mu
m)z\big)\end{array}\right. \eqno(A11)
$$%
We assume that this expansion is valid for all positive values of $z$. We
will show that this solution represents a two-parameter family. The values
of the parameters will come from the boundary conditions (20). Substituting
relations (A11) into (A8) and denoting $\delta =\exp (-\lambda z)$ and $%
\varepsilon =\exp (-\mu z)$, we get
$$
\sum_{k,m\geq 0}(\lambda k+\mu m)^{2}\delta ^{k}\varepsilon
^{m}(m_{xx}x_{km}+m_{xy}y_{km})
$$
$$
=2\sum_{k,m,l,p,q,r\geq 0}x_{km}x_{lp}x_{qr}\delta ^{k+l+q}\varepsilon
^{m+p+r}+2\sum_{k,m,l,p\geq 0}x_{km}y_{lp}\delta ^{k+l}\varepsilon ^{m+p}
$$
and
$$
\sum_{k,m\geq 0}(\lambda k+\mu m)^{2}\delta ^{k}\varepsilon
^{m}(m_{xy}x_{km}+m_{yy}y_{km})
$$
$$
=\sum_{k,m,l,p\geq 0}x_{km}x_{lp}\delta ^{k+l}\varepsilon
^{m+p}+2\sum_{k,m\geq 0}y_{km}\delta ^{k}\varepsilon ^{m}+t\,.
$$
The identification of terms $\delta ^{j}\varepsilon ^{k}$ yields
$$
\delta ^{0}\varepsilon ^{0}:\left(
\begin{array}{c}
x_{00} \\
y_{00}%
\end{array}%
\right) =\left(
\begin{array}{c}
\pm \sqrt{t} \\
-t%
\end{array}%
\right) \ ,
$$
$$
\delta ^{1}\varepsilon ^{0}\ :\left(
\begin{array}{c}
x_{10} \\
y_{10}%
\end{array}%
\right) =c_{10}\mathbf{h}_{10}\mathbf{\ }\text{with }\mathbf{h}_{10}=\left(
\begin{array}{c}
1 \\
0%
\end{array}%
\right) +\mathcal{O}(|x_{00}|)\ ,
$$
$$
\delta ^{0}\varepsilon ^{1}\ :\left(
\begin{array}{c}
x_{01} \\
y_{01}%
\end{array}%
\right) =c_{01}\mathbf{h}_{01}\mathbf{\ }\text{with }\mathbf{h}_{01}=\left(
\begin{array}{c}
-m_{xy} \\
m_{xx}%
\end{array}%
\right) +\mathcal{O}(|x_{00}|)\ .
$$
The vectors $\mathbf{h}_{01}$ and $\mathbf{h}_{10}$ are eigenvectors
corresponding to eigenvalues $\lambda ^{2}$ and $\mu ^{2}$ defined by (A9)
and (A10), respectively.\ The constants $c_{10}$ and $c_{01}$ are multipliers
to be defined.

In the same way, the terms associated with $\delta ^{1}\varepsilon
^{1},\delta ^{0}\varepsilon ^{2}$ and $\delta ^{2}\varepsilon ^{0}$ are
$$
\left(
\begin{array}{c}
x_{11} \\
y_{11}%
\end{array}%
\right) =c_{01}c_{10}\mathbf{h}_{11}\mathbf{\ }\text{with }\mathbf{h}_{11}=-%
\displaystyle\frac{m_{xy}}{D|x_{00}|}\mathbf{h}_{01}+\mathcal{O}(1)\ ,
$$
$$
\left(
\begin{array}{c}
x_{02} \\
y_{02}%
\end{array}%
\right) =c_{01}^{2}\mathbf{h}_{02}\mathbf{\ }\text{with }\mathbf{h}_{02}=%
\frac{m_{xy}}{2}\left(
\begin{array}{c}
\displaystyle\frac{-\left( m_{xx}m_{yy}+m_{xy}^{2}\right) }{2m_{xx}} \\
m_{xy}%
\end{array}%
\right) +\mathcal{O}(|x_{00}|)\ ,
$$
$$
\left(
\begin{array}{c}
x_{20} \\
y_{20}%
\end{array}%
\right) =c_{10}^{2}\mathbf{h}_{20}\mathbf{\ }\text{with }\mathbf{h}%
_{20}=\left(
\begin{array}{c}
\mathcal{O}(|x_{00}|^{-1}) \\
\mathcal{O}(1)%
\end{array}%
\right) \ .
$$

The expansion of $\mathbf{R}$ truncated to the second order with respect to $%
c_{10}$ and $c_{01}$ is
$$
\mathbf{R}=\mathbf{R}_{00}+c_{10}\mathbf{h}_{10}\exp (-\lambda z)+c_{01}%
\mathbf{h}_{01}\exp (-\mu z)+c_{01}c_{10}\mathbf{h}_{11}\exp (-(\lambda +\mu
)z)
$$%
$$
+c_{01}^{2}\mathbf{h}_{02}\exp (-2\mu z)+c_{10}^{2}\mathbf{h}_{20}\exp
(-2\lambda z)+...\eqno(A12)
$$%
Because the solution $\mathbf{R}$ must be bounded as $|x_{00}|$ goes to zero
for all positive values of $z$, $c_{10}$ is at least of order $|x_{00}|$.
Then, we may introduce a constant $\tilde{c}_{10}$ by the formula
$$
c_{10}=D|x_{00}|\tilde{c}_{10},\quad D^{2}=m_{xx}m_{yy}-m_{xy}^{2}\,.
$$%
Since the terms $\mathbf{R}_{00}$ and $c_{10}\mathbf{h}_{10}$ are of the
order of $\mathcal{O}(|x_{00}|)$, we get in the vicinity of $z=0$
$$
\mathbf{R}=a_{01}\mathbf{h}_{01}\exp (-\mu z)+c_{10}^{2}\mathbf{h}_{02}\exp
(-2\mu z)+\mathcal{O}(|x_{00}|),\eqno(A13)
$$%
where all the coefficients are finite as $|x_{00}|$ goes to zero and $%
a_{01}=c_{01}-m_{xy}c_{01}\tilde{c}_{10}$. In fact, the term $\mathbf{R}_{00}
$ is negligible at the vicinity of the wall, but not in the bulk.

\noindent For the sake of simplicity, we exhibit the boundary conditions
only in the case of a purely attractive wall (when $\Gamma$ is negligible).%
\newline
Then, the condition (20) on the wall reads
$$
M{\frac{d\mathbf{R}}{dz}} = - \Delta^{ \star}{\mathbf{\gamma}} \quad \mathrm{%
at}\quad z = 0.
$$
Or, by using (A13), we get
$$
\mu_0 M( {a}_{01} \mathbf{h}_{01} + 2 c_{10}^2 \mathbf{h}_{02}) = \Delta^{
\star}{\mathbf{\gamma}} \,.\eqno(A14)
$$
By multiplying (A14) by $\mathbf{h}_{10}$ and $\mathbf{h}_{01}$, we get a
system of two scalar equations. By taking into account the equality $\mathbf{%
h}_{10} M \mathbf{h}_{01} = 0 $, the vector equation (A14) is equivalent to
the system of two scalar equations
$$
2 \mu_0 c_{10}^2 \mathbf{h}_{10} M \mathbf{h}_{02} = \mathbf{h}_{10}
\Delta^{ \star}{\mathbf{\gamma}} \eqno(A15)
$$
$$
\mu_0 \left(a_{01} \mathbf{h}_{01} M \mathbf{h}_{01} + 2 c_{10}^2 \mathbf{h}%
_{01} M \mathbf{h}_{02} \right) = \mathbf{h}_{01} \Delta^{ \star}{\mathbf{%
\gamma}}\,. \eqno(A16)
$$
Eq (A15) defines $c_{10}^2$ and then, Eq. (A16) defines $a_{01}$. Hence the
solution near the wall (A13) is completely determined. Consequently, we are
able to determine the effect of the solid wall on the adsorption of each
component. Due to the fact that the expansion (A13) depends only on ${c_{01}}%
^2$ and not $c_{01}$, in this approximation we do not need to satisfy the
inequality ${c_{01}}^2 > 0$.

\bigskip

\noindent \textbf{References}

\noindent [1] J. Cahn, \textit{J. Chem. Phys.} \textbf{66} (1977) 3667.
\newline
\noindent [2] D.E. Sullivan, \textit{Phys. Rev. B} \textbf{20} (1979) 3991.
\newline
\noindent [3] D.E. Sullivan, \textit{J. Chem. Phys.} \textbf{74} (1981)
2604. \newline
\noindent [4] M. Telo da Gama and R. Evans, \textit{Mol. Phys.} \textbf{48}
(1983) 687. \newline
\noindent [5] I. Hadjiagapiou and R. Evans, \textit{Mol. Phys.} \textbf{54}
(1985) 283. \newline
\noindent [6] H. Gouin, \textit{J. Phys. Chem. B} \textbf{102} (1998) 1212.
\newline
\noindent [7] J.S. Rowlinson and B. Widom, \textit{Molecular Theory of
Capillarity}, Clarendon Press, Oxford, 1984. \newline
\noindent [8] N.G. van Kampen, \textit{Phys. Rev. B} \textbf{135} (1964) A
362. \newline
\noindent [9] B. Widom, \textit{J. Stat. Phys.} \textbf{19} (1978) 563.
\newline
\noindent [10] P.D. Fleming, A.J.M. Yang, J.H. Gibbs, \textit{J. Chem. Phys.}
\textbf{65}, (1976) 7. \newline
\noindent [11] H. Nakanishi and M.E. Fisher \textit{Phys. Rev. Lett.}
\textbf{49} (1982) 1565. \newline
\noindent [12] P.G. de Gennes, \textit{Rev. Mod. Phys.} \textbf{57} (1985)
827. \newline
\noindent [13] H. Gouin, \textit{Eur. J. Mech. B/ Fluids} \textbf{9} (1990)
469. \newline
\noindent [14] J. Serrin, \textit{Mathematical principles of classical fluid
mechanics}, Encyclopedia of Physics, VIII/1, Springer, Berlin, 1959. pp.
125-263. \newline
\noindent[15] S.L. Gavrilyuk, H. Gouin, Yu. V. Perepechko, \textit{C. R.
Acad. Sci. Paris} \textbf{II 324} (1997) 483.

\end{document}